# Degeneracy: a link between evolvability, robustness and complexity in biological systems

James M Whitacre

*School of Computer Science, University of Birmingham, Edgbaston, UK*

## 1.1.    Abstract

A full accounting of biological robustness remains elusive; both in terms of the mechanisms by which robustness is achieved and the forces that have caused robustness to grow over evolutionary time.  Although its importance to topics such as ecosystem services and resilience is well recognized, the broader relationship between robustness and evolution is only starting to be fully appreciated.  A renewed interest in this relationship has been prompted by evidence that mutational robustness can play a positive role in the discovery of future adaptive innovations (evolvability) and evidence of an intimate relationship between robustness and complexity in biology.

This paper offers a new perspective on the mechanics of evolution and the origins of complexity, robustness, and evolvability.  Here we explore the hypothesis that degeneracy, a partial overlap in the functioning of multi-functional components, plays a central role in the evolution and robustness of complex forms. In support of this hypothesis, we present evidence that degeneracy is a fundamental source of robustness, it is intimately tied to multi-scaled complexity, and it establishes conditions that are necessary for system evolvability.

# 1. Introduction

Complex adaptive systems (CAS) are omnipresent and are at the core of some of society's most challenging and rewarding endeavours. They are also of interest in their own right because of the unique features they exhibit such as high complexity, robustness, and the capacity to innovate.  Especially within biological contexts such as the immune system, the brain, and gene regulation, CAS are extraordinarily robust to variation in both internal and external conditions.  This robustness is in many ways unique because it is conferred through rich distributed responses that allow these systems to handle challenging and varied environmental stresses.  Although exceptionally robust, biological systems can sometimes adapt in ways that exploit new resources or allow them to persist under unprecedented environmental regime shifts.

These requirements to be both robust and adaptive appear to be conflicting. For instance, it is not entirely understood how organisms can be phenotypically robust to genetic mutations yet also can generate the range of phenotypic variability that is needed for evolutionary adaptations to occur. Moreover, on rare occasions genetic changes can result in increased system complexity however it is not known how these increasingly complex forms are able to evolve without sacrificing robustness or the propensity for future beneficial adaptations. To put it more distinctly, it is not known how biological evolution is scalable [1].

A deeper understanding of CAS thus requires a deeper understanding of the conditions that facilitate the coexistence of high robustness, growing complexity, and the continued propensity for innovation or what we refer to as evolvability. This reconciliation is not only of interest to biological evolution but also to science in general because variability in conditions and unprecedented shocks are a challenge faced across many facets of human enterprise.

In this opinion paper, we explore and expand upon the hypothesis first proposed in [2] [3] that a system property known as degeneracy plays a central role in the relationships between these properties. Most importantly, we argue that only robustness through degeneracy will lead to evolvability or to hierarchical complexity in CAS.  An overview of our main arguments is shown in Figure 1 with Table 1 summarizing primary supporting evidence from the literature. Throughout this paper, we refer back to Figure 1  so as to connect individual discussions with the broader hypothesis being proposed. For instance, we refer to "Link 6" in the heading of Section 2 in reference to the connection between robustness and evolvability that is to be discussed and also that is shown as the sixth link in the graph in Figure 1.

The remainder of the paper is organized as follows. We begin by reviewing the paradoxical relationship between robustness and evolvability in biological evolution.  Starting with evidence that robustness and evolvability can coexist, in Section 2 we present arguments for why this is not always the case in other domains and how degeneracy might play an important role in reconciling these conflicting properties. Section 3 outlines further evidence that degeneracy is causally intertwined within the unique relationships between robustness, complexity, and evolvability in CAS.  We discuss its prevalence in biological systems, its role in establishing robust traits, and its relationship with information theoretic measures of hierarchical complexity. Motivated by these discussions, we speculate in Section 4 that degeneracy may provide a mechanistic explanation for the theory of natural selection and particularly some more recent hypotheses such as the theory of highly optimized tolerance.

# 2. Robustness and Evolvability (Link 6)

Phenotypic robustness and evolvability are defining properties of CAS. In biology, the term robustness is often used in reference to the persistence of high level traits, e.g. fitness, under variable conditions. In contrast, evolvability refers to the capacity for heritable and selectable phenotypic change. More thorough descriptions of robustness and evolvability can be found in Box1.

Robustness and evolvability are vital to the persistence of life and their relationship is vital to our understanding of it. This is emphasized in [4] where Wagner asserts that, "*understanding the relationship between robustness and evolvability is key to understand how living things can withstand mutations, while producing ample variation that leads to evolutionary innovations*". At first, robustness and evolvability appear to be in conflict as suggested in the study of RNA secondary structure evolution by Ancel and Fontana [5]. As an illustration of this conflict, the first two panels in Figure 2 show how high phenotypic robustness appears to imply a low production of heritable phenotypic variation [4]. These graphs reflect common intuition that maintaining developed functionalities while at the same time exploring and finding new ones are contradictory requirements of evolution.

## *2.1. Resolving the robustness-evolvability conflict*

However, as demonstrated in [4] and illustrated in panel c of Figure 2, this conflict is unresolvable only when robustness is conferred in both the genotype and the phenotype. On the other hand, if the phenotype is robustly maintained in the presence of genetic mutations, then a number of cryptic genetic changes may be possible and their accumulation over time might expose a broad range of distinct phenotypes, e.g. by movement across a neutral network. In this way, robustness of the phenotype might actually enhance access to heritable phenotypic variation and thereby improve long-term evolvability.

The work by Ciliberti et al [6] represents a useful case study for understanding this resolution of the robustness/evolvability conflict, although we note that earlier studies arguably demonstrated similar phenomena [7] [8]. In [6], the authors use models of gene regulatory networks (GRN) where GRN instances represent points in genotype space and their expression pattern represents an output or phenotype. Together the genotype and phenotype define a fitness landscape. With this model, Ciliberti et al find that a large number of genotypic changes to the GRN have no phenotypic effect, thereby indicating robustness to such changes. These phenotypically equivalent systems connect to form a neutral network NN in the fitness landscape. A search over this NN is able to reach nodes whose genotypes are almost as different from one another as randomly sampled GRNs. The authors also find that the number of distinct phenotypes that are in the local vicinity of NN nodes is extremely large, indicating a wide variety of accessible phenotypes that can be explored while remaining close to a viable phenotype. The types of phenotypes that are accessible from the NN depend on where in the network that the search takes place. This is evidence that cryptic genetic changes (along the NN) eventually have distinctive phenotypic consequences.

In short, the study presented in [6] suggests that the conflict between robustness and evolvability is resolved through the existence of a NN that extends far throughout the fitness landscape. On the one hand, robustness is achieved through a connected network of equivalent (or nearly equivalent) phenotypes. Because of this connectivity, some mutations or perturbations will leave the phenotype unchanged, the extent of which depends on the local NN topology. On the other hand, evolvability is achieved over the long-term by movement across a neutral network that reaches over truly unique regions of the fitness landscape.

## Robustness and evolvability are not always compatible

A positive correlation between robustness and evolvability is widely believed to be conditional upon several other factors, however it is not yet clear what those factors are. Some insights into this problem can be gained by comparing and contrasting systems in which robustness is and is not compatible with evolvability.

In accordance with universal Darwinism [9], there are numerous contexts where heritable variation and selection take place and where evolutionary concepts can be successfully applied. These include networked technologies, culture, language, knowledge, music, markets, and organizations. Although a rigorous analysis of robustness and evolvability has not been attempted within any of these domains, there is anecdotal evidence that evolvability does not always go hand in hand with robustness. Many technological and social systems have been intentionally designed to enhance the robustness of a particular service or function, however they are often not readily adaptable to change. In engineering design in particular, it is a well known heuristic that increasing robustness and complexity can often be a deterrent to flexibility and future adaptations. Similar trade-offs surface in the context of governance (bureaucracy), software design (e.g operating systems), and planning under high uncertainty (e.g. strategic planning).

Other evidence of a conflict between robustness and evolvability has been observed in computer simulations of evolution. Studies within the fields of evolutionary computation and artificial life have considered ways of manually injecting mutational robustness into the mapping of genotype to phenotype, e.g. via the enlargement of neutral regions within fitness landscapes [10] [11] [12] [13] [14]. Adding mutational robustness in this way has had little influence on the evolvability of simulated populations. Some researchers have concluded that genetic neutrality (i.e. mutational robustness) alone is not sufficient. Instead, it has been argued that the positioning of neutrality within a fitness landscape through the interactions between genes will greatly influence the number and variety of accessible phenotypes [15] [16].

Assessing the different domains where variation and selection take place, it is noticeable that evolvability and robustness are often in conflict within systems derived through human planning. But how could the simple act of planning change the relationship between robustness and evolvability? As first proposed by Edelman and Gally, one important difference between systems that are created by design (i.e. through planning) and those that evolve without planning is that in the former, components with multiple overlapping functions are absent [2].

In standard planning practices, components remain as simple as possible with a single predetermined functionality. Irrelevant interactions and overlapping functions between components are eliminated from the outset, thereby allowing cause and effect to be more transparent. Robustness is achieved by designing redundancies into a system that are predictable and globally controllable [2].

This can be contrasted with biological CAS such as gene regulatory networks or neural networks where the relevance of interactions can not be determined by local inspection. There is no predetermined assignment of responsibilities for functions or system traits. Instead, different components can contribute to the same function and a component can contribute to several different functions through its exposure to different contexts. While the functionalities of some components appear to be similar under specific conditions, they differ under others. This conditional similarity of functions within biological CAS is a reflection of degeneracy.

# 3. Degeneracy

Degeneracy is a system property that requires the existence of multi-functional components (but also modules and pathways) that perform similar functions (i.e. are effectively interchangeable) under certain conditions, yet can perform distinct functions under other conditions. A case in point is the adhesins gene family in *A. Saccharomyces*, which expresses proteins that typically play unique roles during development, yet can perform each other's functions when expression levels are altered [17]. Another classic example of degeneracy is found in glucose metabolism, which can take place through two distinct pathways; glycolysis and the pentose phosphate pathway. These pathways can substitute for each other if necessary even though the sum of their metabolic effects is not identical [18]. More generally, Ma and Zeng argue that the robustness of the bow-tie architecture they discovered in metabolism is largely derived through the presence of multiple degenerate paths to achieving a given function or activity [19] [20]. Although we could list many more examples of degeneracy, a true appreciation for the ubiquity of degeneracy across all scales of biology is best gained by reading Edelman and Gally's review of the topic in [2]. Box 2 provides a more detailed description of degeneracy, its relationship to redundancy, and additional examples of degeneracy in biological systems.

### 3.1.    The role of degeneracy in adaptive innovations  (Links 1 & 3)

In [3], we explored whether degeneracy influences the relationship between robustness and evolvability in a generic genome:proteome model. Unlike the studies discussed in Section 2, we found that neither size nor topology of a neutral network guarantees evolvability.  Local and global measures of robustness within a fitness landscape were also not consistently indicative of the accessibility of distinct heritable phenotypes. Instead, we found that only systems with high levels of degeneracy exhibited a positive relationship between neutral network size, robustness, and evolvability.

More precisely, we showed that systems composed of redundant proteins were mutationally robust but greatly restricted in the number of unique phenotypes accessible from a neutral network, i.e. they were not evolvable.  On the other hand, replacing redundant proteins with degenerate proteins resolved this conflict and led to both exceptionally robust and exceptionally evolvable systems.  Importantly, this result was observed even though the total sum of protein functions was identical between each of the system classes. From observing how evolvability scaled with system size, we concluded that degeneracy not only contributes to the discovery of new innovations but that it may be a precondition of evolvability [21] [3].

## Degeneracy and distributed robustness (Link 1)

As discussed in [2], degeneracy's relationship to robustness and evolvability appears to be conceptually simple. While degenerate components contribute to stability under conditions where they are functionally compensatory, their distinct responses outside of those conditions provide access to unique functional effects, some of which may be selectively relevant in certain environments.

Although useful in guiding our intuition, it is not clear whether such ideas are applicable to larger systems involving many components and multiple traits. More precisely, it is not clear why accessibility to functional variation would not act as a destabilizing force within a larger system. However in [3], we found that the robustness  of large degenerate

genome:proteome systems was not degraded by this functional variation and instead was actually greater than that expected from local compensatory actions. In the following, we present an alternative conceptual model that aims to account for these findings and illustrates additional ways in which degeneracy may facilitate robustness and evolvability in complex adaptive systems.

The model comprises agents that are situated within an environment. Each agent can perform one task at a time where the types of tasks are restricted by an agent's predetermined capabilities. Tasks represent conditions imposed by the local environment and agents act to take on any tasks that match their functional repertoire. An illustration of how degeneracy can influence robustness and evolvability is given using the diagrams in Figure 3, where each task type is represented by a node cluster and agents are represented by pairs of connected nodes. For instance, in Figure 3a an agent is circled and the positioning of its nodes reflects that agent's (two) task capabilities. Each agent only performs one task at a time with the currently executed task indicated by the darker node.

In Figure 3b, task requirements are increased for the bottom task group and excess resources become available in the top task group. With a partial overlap in agent task capabilities, agent resources can be reassigned from where they are in excess to where they are needed as indicated by the arrows. From this simple illustration, it is straightforward to see how excess buffering related to one type of task can support unrelated tasks through the presence of degeneracy. If this partial overlap in capabilities is pervasive throughout the system then there are potentially many options for reconfiguring resources as shown in Figure 3c. In short, degeneracy may allow for cooperation amongst buffers such that localized stresses can invoke a distributed response. Moreover, excess resources related to a single task can be used in a highly versatile manner; although interoperability of components may be localized, at the system level extra resources can offer huge reconfiguration opportunities.

The necessary conditions for this buffering network to form do not appear to be demanding (e.g. [3]). One condition that is clearly needed though is degeneracy. Without a partial overlap in capabilities, agents in the same functional grouping can only support each other (see Figure 3d) and, conversely, excess resources cannot support unrelated tasks outside the group. Buffers are thus localized and every type of variability in task requirements requires a matching realization of redundancies. This simplicity in structure (and inefficiency) is encouraged in most human planning activities.

## Degeneracy and Evolvability (Link 3)

For systems to be both robust and evolvable, the individual agents that stabilize traits must be able to occasionally behave in unique ways when stability is lost. Within the context of distributed genetic systems, this requirement is reflected in the need for unique phenotypes to be mutationally accessible from different regions of a neutral network.

The large number of distinct and cryptic internal configurations that are possible within degenerate systems (see Figure 3c) are likely to expand the number of unique ways in which a system will reorganize itself when thresholds for trait stability are eventually crossed, as seen in [3]. This is because degenerate pathways to robust traits are reached by truly distinct paths (i.e. distinct internal configurations) that do not always respond to environmental changes in the same manner, i.e. they are only conditionally similar. Due to symmetry, such cryptic distinctions are not possible from purely redundant sources of robustness.

However, in [3] degenerate systems were found to have an elevated configurational versatility that we speculate is the result of degenerate components being organized into a larger buffering network. This versatility allows degenerate components to contribute to the mutational robustness of a large heterogeneous system and, for the same (symmetry) reasons as stated above, may further contribute to the accessibility of distinct heritable variation.

In summary, we have presented arguments as well as some evidence that degeneracy allows for types of robustness that directly contribute to the evolvability of complex systems, e.g. through mutational access to distinct phenotypes from a neutral network within a fitness landscape. We have speculated that the basis for both robustness and evolvability in degenerate systems is a set of heterogeneous overlapping buffers. We suggest that these buffers and their connectivity offer exceptional canalization potential under many conditions while facilitating high levels of phenotypic plasticity under others.

# 4. Origins of complexity

## 4.1. Complexity

There are many definitions and studies of complexity in the literature [22] [23] [24] [25] [26] [27] [28]. Different definitions have mostly originated within separate disciplines and have been shaped by the classes of systems that are considered pertinent to particular fields of study.

Early usage of the term complexity within biology was fairly ambiguous and varied depending on the context in which it was used. Darwin appeared to equate complexity with the number of distinct components (e.g. cells) that were "organized" to generate a particular trait (e.g. an eye). Since then, the meaning of complexity has changed however nowadays only marginal consensus exists on what it means and how it should be measured. In studies related to the theory of highly optimized tolerance (HOT), complex systems have been defined as being hierarchical, highly structured and composed of many heterogeneous components [29] [30].

The organizational structure of life is now known to be scale-rich (as opposed to scale-free) but also multi-scaled [31] [29] [30]. This means that patterns of biological component interdependence are truly unique to a particular scale of observation but there are also important interactions that integrate behaviors across scales.

The existence of expanding hierarchical structures or "systems within systems" implies a scalability in natural evolution that some would label as a uniquely biological phenomenon. From prions and viruses to rich ecosystems and the biosphere, we observe organized systems that rely heavily on the robustness of finer-scale patterns while they also adapt to change taking place at a larger scale [32].

A defining characteristic of multi-scaled complex systems is captured in the definition of hierarchical complexity given in [33] [34]. There, complexity is defined as the degree to which a system is both functionally integrated and functionally segregated. Although this may not express what complexity means to all people, we focus on this definition because it represents an important quantifiable property of multi-scaled complex systems that is arguably unique to biological evolution.

## 4.2.     Degeneracy and Complexity (Link 2)

According to Tononi et al [33], degeneracy is intimately related to complexity, both conceptually as well as empirically. The conceptual similarity is immediately apparent: while complex systems are both functionally integrated and functionally segregated, degenerate components are both functionally redundant and functionally independent. Tononi et al also found that a strong positive correlation exists between information theoretic measurements of degeneracy and complexity. When degeneracy was increased within neural network models, they always observed a concomitant large increase in system complexity. In contrast, complexity was found to be low in cases where neurons fired independently (although Shannon entropy is high in this case) or when firing throughout the neuronal population was strongly correlated (although information redundancy is high in this case). From these observations, Tononi et al derived a more generic claim, namely that this relationship between degeneracy and complexity is broadly relevant and could be pertinent to our general understanding of CAS.

## 4.3.     Robustness and Complexity (Link 5)

System robustness requires that components can be "utilized" at the appropriate times to accommodate aberrant variations in the conditions to which a system is exposed. Because such irregular variability can be large in both scale and type, robustness is limited by the capabilities of extant components. Such limitations are easily recognizable and commonly relate to limits on utilization rate and level of multi-functionality afforded to any single component. As a result of these physical constraints, improvements in robustness can sometimes only occur from the integration of new components and new component types within a system, which in turn can add to a system's complexity.

While the integration of new components may address certain aberrant variations in conditions, it can also introduce new degrees of freedom to the system which sometimes leads to new points of accessible fragility, i.e. new vulnerabilities. As long as the frequency and impact of conditions stabilized is larger than those of the conditions sensitised, such components are theoretically selectable by robustness measures.[1] By this reasoning, a sustained drive towards increased robustness might be expected to correspond occasionally with growth in system complexity. At sufficiently long time scales, we thus might expect a strong positive correlation to emerge between the two properties.

Such a relationship between robustness and complexity is proposed in the theory of Highly Optimized Tolerance (HOT) [29] [35] [30]. HOT suggests that the myopic nature of evolution promotes increased robustness to common conditions and unknowingly replaces these with considerably less frequent but still potentially devastating sensitivities. Proponents of HOT argue that evidence of this process can be found in the properties of evolving systems, such as power law relations observed for certain spatial and temporal properties of evolution, e.g. extinction sizes. In support of HOT, some researchers have used examples from biology, ecology, and engineering to demonstrate how increased robustness often simultaneously leads to increased system complexity [29] [35] [30].

From another perspective, the persistence of a heterogeneous, multi-scaled system seems to necessitate robustness, at least to intrinsic variability that may arise, for instance, from process errors initiated by the stochasticity of internal dynamics. Without such robustness, small aberrant perturbations in one "subsystem" could spread to others, leading to broad destabilization of these subsystems and a potential collapse of otherwise persistent higher-scale patterns. Even if individual perturbations are unlikely, the frequency of perturbation events (e.g. at fine resolutions of the system) would greatly limit the overall number of distinct scales where coherent spatio-temporal patterns could be observed, if the system were not robust. Similar arguments have been used in explaining the relationship between multi-scaling phenomena and resilience within complex ecosystems [36] [32] [37].

---

[1] This makes assumptions about the timescales of variation and selection. For instance, it assumes that the environment is not changing "too rapidly". If current conditions do not reflect future conditions (to at least "some degree") then any increases in robustness could only occur by chance and generally would not be observed.

**The role of degeneracy:** Summarizing, it is apparent that robustness and complexity are intimately intertwined and moreover that robustness is a precondition for complexity, at least for multi-scaled systems. However, not all mechanisms for achieving robustness necessarily lead to multi-scaled complexity nor are they all viable options for CAS. For instance, in [33] Tononi et al found that highly redundant (non-degenerate) systems were naturally robust but never hierarchically complex. On the other hand, highly degenerate systems were simultaneously robust and complex. Assuming as Tononi et al do that their findings extend to other CAS, this suggests that the relationship between robustness and complexity hypothesized in HOT is enabled by the presence of degenerate forms of robustness.

### 4.4.    Evolution of complex phenotypes (Link 4)

The evolution of complex forms requires a long series of adaptive changes to take place. At each step, these adaptations must result in a viable and robust system but also must not inhibit the ability to find subsequent adaptations. Complexity, in the context of multi-scaled evolving systems, clearly demands evolvability to form such systems and robustness to maintain such systems at every step along the way. This connection between evolvability and complexity is famously captured within Darwin's theory of natural selection. According to the theory, complex traits have evolved through a series of incremental changes and are not irreducibly complex. For highly complex traits to exist, growth in complexity cannot inhibit the future evolvability of a system. More precisely, the formation of complex traits is predicated on evolvability either being sustained or re-emerging after each inherited change.

How evolving systems actually satisfy these requirements remains a true mystery. As reviewed by Kirschner and Gerhart, different principles in biological systems have been uncovered over the years (e.g. loose-coupling, exploratory behavior, redundancy, agent versatility) that strongly influence the constraint/deconstraint mechanisms imposed on phenotypic variation and thus contribute to robustness and evolvability of these systems [1]. Although these principles are undoubtedly of considerable importance, the examples of degeneracy provided in [2] strongly suggest that degeneracy underpins most constraint/deconstraint mechanisms.

It is well-accepted that the exceptional properties of CAS are not a consequence of exceptional properties of their components [23]. Instead it is how components interact and inter-relate that determines: 1) the ability to confer stability within the broader system (robustness), 2) the ability to create systems that are both functionally integrated and functionally segregated (complex), and 3) the ability to acquire new traits and take on more complex forms (evolvable). It would seem that any mechanism that directly contributes to all of these organizational properties is a promising candidate design principle of evolution. In this paper we have reviewed new evidence, summarized in Table 1 and Figure 1, that degeneracy may represent just such a mechanism and thus could prove fundamental in understanding the evolution of complex forms. But if degeneracy is important to the mechanics of evolution as claimed in this paper, it is worth asking why it has been overlooked in theoretical discussions of biological evolution. As suggested in Box 3, we argue that this may be due to a long-standing reductionist bias in the study of biological systems.

# 5. Concluding Remarks

Understanding how biological systems can be complex, robust and evolvable is germane to our understanding of biology and evolution. In this paper, we have proposed that degeneracy could play a fundamental role in the unique relationships between complexity, robustness, and evolvability in complex adaptive systems. Summarizing our arguments, we have presented evidence that degeneracy is a highly effective mechanism for creating (distributed) robust systems, that it uniquely enhances the long-term evolvability of a system, that it acts to increase the hierarchical complexity of a system, and that it is prevalent in biology. Using these arguments, we speculate on how degeneracy may help to directly establish the conditions necessary for the evolution of complex forms. Although more research is needed to validate some of the claims made here, we are cautiously optimistic that degeneracy is intimately tied to some of the most interesting phenomena observed in natural evolving systems. Moreover, as a conceptual design principle, degeneracy is readily applicable to other disciplines and could prove beneficial for enhancing the robustness and adaptiveness of human-engineered systems.

# Box 1:  Robustness and Evolvability

In nature, organisms are presented with a multitude of environments and are occasionally exposed to new and slightly different environments. Under these variable conditions, organisms must on the one hand maintain a range of functionalities in order to survive and reproduce. Often, this means a number of important traits need to be robust over a range of environments. On the other hand, organisms must also be flexible enough to adapt to new conditions that they have not previously experienced. At higher levels in biology, populations display genetic robustness and robustness to moderate ecological changes yet at the same time are often able to adapt when conditions change "significantly". This dual presence of robustness and adaptiveness to change is observed at different scales in biology and it has been responsible for the remarkable persistence of life over billions of years and countless generations.

**Robustness**

Despite the numerous definitions of robustness provided in the literature [38], there is fair conceptual agreement on what robustness means. In its most general form, robustness reflects an insensitivity of some functionality or measured state of a system when the system is exposed to a set of distinct environments or distinct internal conditions. To give robustness meaning, it is necessary to elaborate on what function or state of the system is being measured and to what set of conditions the system is exposed.

**Classes of Environmental and Biological Change:** The conditions to which a system is exposed depend on its scale and scope but are generally broken down into internal and external sources. For instance, changes originating from within an organism include inherited changes to the genotype and stochasticity of internal dynamics, while sources of external (environmental) change include changes in culture, changes in species interactions and changes at various scales within the physical environment.

**Pathways toward robustness**
Biological robustness is typically discussed as a process of effective control over the phenotype. In some cases, this means maintaining a stable trait despite variability in the environment (canalization), while in other cases it requires modification of a trait so as to maintain higher level traits such as fitness, within a new environment (adaptive phenotypic plasticity) [19]. Both adaptive phenotypic plasticity and canalization involve conditional responses to change and their causal origins are generally believed to be similar [39].

# Evolvability

Different definitions of evolvability exist in the literature (e.g. [4] [40] [41]), so it is important to articulate exactly what is meant by this term. In general, evolvability is concerned with the selection of new phenotypes. It requires an ability to generate distinct phenotypes and it requires that some of these phenotypes have a non-negligible probability of being selected by the environment.

Because of the contextual nature of selection (i.e. its dependence on the environment), quantifying evolvability in a "context free" manner is only possible by employing a surrogate measurement. The most common measurement used in the literature is the accessibility of distinct heritable phenotypes [1]. In this paper, as with others [6] [4] [42], we use this surrogate measure when evaluating evolvability.

# Box 2: Degeneracy and Redundancy

Redundancy and degeneracy are two design principles that both contribute to the robustness of biological systems [2] [43]. Redundancy is an easily recognizable design principle in biological and man-made systems and means 'redundancy of parts'. It refers to the coexistence of identical components with identical functionality and thus is isomorphic and isofunctional. In information theory, redundancy refers to the repetition of messages and is important for reducing transmission errors. It is a common feature of engineered or planned systems where it provides robustness against variations of a very specific type ('more of the same' variations). For example, redundant parts can substitute for others that malfunction or fail, or augment output when demand for a particular output increases. Redundancy is also prevalent in biology. Polyploidy, homogenous tissues and allozymes are examples of functional biological redundancy. Another and particular impressive example is neural redundancy, i.e. the multiplicity of neural units (e.g. pacemaker cells) that perform identical functions (e.g. generating the swimming rhythms in jellyfish or the heartbeat in humans).

In biology, degeneracy refers to conditions where the functions or capabilities of components overlap partially. In a review by Edelman and Gally [2], numerous examples are used to demonstrate the ubiquity of degeneracy throughout biology. It is pervasive in proteins of every functional class (e.g. enzymatic, structural, or regulatory) [44] and is readily observed in ontogenesis (see page 14 in [45]), the nervous system [33] and cell signalling (crosstalk). Degeneracy differs from pure redundancy because similarities in the functional response of components are not observed for all conditions. Under some conditions the functions are similar while under others they differ.

**Origins of Degeneracy**
Degeneracy originates from convergent (constraint) and divergent (deconstraint) forces that play out within distributed systems subject to variation and selection. With divergence, identical components evolve in slightly distinct directions causing structural and functional differences to grow over time. The most well studied context where this occurs is gene duplication and divergence [46] [47] [48]. Degeneracy may also arise through convergent evolution, where structurally distinct components are driven to acquire similar functionalities. In biology, this may occur as a direct result of selection for a particular trait or it may alternatively arise due to developmental constraints (e.g. see [49]) that act to constrain the evolution of dissimilar components in similar ways. There are many documented examples of convergence (e.g. homoplasy) occurring at different scales in biology [50] [51].

While the origins of degeneracy are conceptually simple, the reasons it is observed at high levels throughout biology are not known and several plausible explanations exist. One possibility is that degeneracy is expressed at high levels simply because it is the quickest or most probable path to heritable change in distributed (genetic) systems. Another possibility is that it is retained due to a direct selective advantage, e.g. due to the enhanced robustness it may provide towards variability in the environment, e.g. see [3]. Other interesting explanations have been proposed that consider a combination of neutral and selective processes. For instance, the Duplication-Degeneracy-Complementation (DDC) model [52] proposes that neutral drift can readily introduce degeneracy amongst initially redundant genes that is later fixated through complementary loss-of-function mutations. Yet another possibility proposed in [3] is that the distributed nature of degenerate robustness (e.g. see Figure 3c) creates a relatively large mutational target for trait buffering that is separate from the degenerate gene. This large target may help to increase and preserve degeneracy over iterated periods of addition and removal of excess robustness within populations under mutation-selection balance (cf [3]). Similar to the DDC model, under this scenario degeneracy would be acquired passively (neutrally) and selectively retained only after additional loss-of-function mutations.

# Box 3: The "hidden" role of degeneracy

If degeneracy is important to the mechanics of evolution as claimed in this paper, it is worth asking why it has been overlooked in theoretical discussions of biological evolution. In [2], Edelman and Gally suggest that its importance has been hidden in plain sight but that the ubiquity of degeneracy and its importance to evolutionary mechanics become obvious upon close inspection. We believe there may also be practical reasons degeneracy has been overlooked which originate from a long-standing reductionist bias in the study of biological systems.

An illustrative example is given by the proposed relationship between degeneracy and robustness. As speculated in Figure 3, degeneracy contributes to robustness through distributed compensatory actions whereby: i) distinct components support the stability of a single trait and ii) individual components contribute to the stability of multiple distinct traits. However, the experimental conditions of biological studies are rarely designed to evaluate emergent and systemic causes of trait stability. Instead, biological studies often evaluate single trait stability or only evaluate mechanisms that stabilize traits through local interactions, e.g. via functional redundancy in a single specified context. This is evident within the many studies and examples of trait stability reviewed in [2].

Degeneracy's influence on evolvability is also largely hidden when viewed from a reductionist lens. As already discussed, the (internal) organizational versatility afforded by degeneracy can allow many perturbations to have a neutral or muted phenotypic effect. When phenotypic innovations do eventually occur however, they are likely to be influenced by the many cryptic changes occurring prior to the final threshold crossing event, e.g. mutation [53] [54] [55]. While the single gene:trait paradigm has long been put to rest, studies investigating phenotypic variation still often rely on single gene knockout experiments and simple models of gene epistasis. Historically, studies have rarely been designed in a manner that could expose the utility of neutral/passive mechanistic processes in facilitating adaptive change [53].

Degenerate components often have many context-activated functional effects and frequent changes to context can cause a component's influence to be highly variable over time. The prevalence of spatio-temporal variability in function has been well documented in the proteome where the most versatile of such proteins are labelled as date-hubs [56]. However, most biological data sets are obtained using time-averaged measurements of effect size which can make versatile components appear to have weak interactions even though these interactions are relevant to trait stability. This limitation from time-averaged measurement bias was first demonstrated by Berlow for species interactions within intertidal ecological communities [57]. However, even if highly versatile components do exhibit a relatively low affinity in each of their interactions, they may still have a large influence on system coherence, integration, and stability [58] [59]. For instance, the low affinity "weak links" of some degenerate components are known to play a vital role in the stability of social networks [60] and within the cell's interactome, e.g. protein chaperones [59]. However, for reasons associated with time and cost restrictions, weak links are typically discounted in both data collection and analysis of biological systems. In summary, we suspect that commonly accepted forms of experimental bias and conceptual (reductionist) bias have hindered scientific exploration of degeneracy and it's role in facilitating phenotypic robustness and evolvability.

# 6. Tables

**Table 1 Overview of key studies on the relationship between degeneracy, robustness, complexity and evolvability. The information is mostly taken (with permission) from [3].**

| | Relationship | Summary | Context | Ref |
|---|---|---|---|---|
| 1) | Unknown whether degeneracy is a primary source of robustness in biology | Distributed robustness (and not pure redundancy) accounts for a large proportion of robustness in biological systems (Kitami, 2002), (Wagner, 2005). Although many traits are stabilized through degeneracy (Edelman and Gally, 2001) its total contribution is unknown. | Large scale gene deletion studies and other biological evidence (e.g. cryptic genetic variation) | [43], [61] [2] |
| 2) | Degeneracy has a strong positive correlation with system complexity | Degeneracy is positively correlated and conceptually similar to complexity. For instance degenerate components are both functionally redundant and functionally independent while complexity describes systems that are functionally integrated and functionally segregated. | Simulation models of artificial neural networks are evaluated based on information theoretic measures of redundancy, degeneracy, and complexity | [33] |
| 3) | Degeneracy is a precondition for | Accessibility of distinct phenotypes requires robustness through degeneracy | Abstract simulation models of evolution | [3] |
| | evolvability and a more effective source of robustness | | | |
| 4) | Evolvability is a prerequisite for complexity | All complex life forms have evolved through a succession of incremental changes and are not irreducibly complex (according to Darwin's theory of natural selection). The capacity to generate heritable phenotypic variation (evolvability) is a precondition for the evolution of increasingly complex forms. | Theory of natural selection | [62] |
| 5) | Complexity increases to improve robustness | According to the theory of highly optimized tolerance, complex adaptive systems are optimized for robustness to common observed variations in conditions. Moreover, robustness is improved through the addition of new components/processes that are integrated with the rest of the system and add to the complexity of the organizational form. | Based on theoretical arguments that have been applied to biological evolution and engineering design (e.g. aircraft, internet) | [29] [35] [30] |
| 6) | Evolvability emerges from robustness | Genetic robustness reflects the presence of a neutral network. Over the long-term this neutral network provides access to a broad range of distinct phenotypes and helps ensure the long-term evolvability of a system. | Simulation models of gene regulatory networks and RNA secondary structure. | [6] [4] |

# 7. Figures

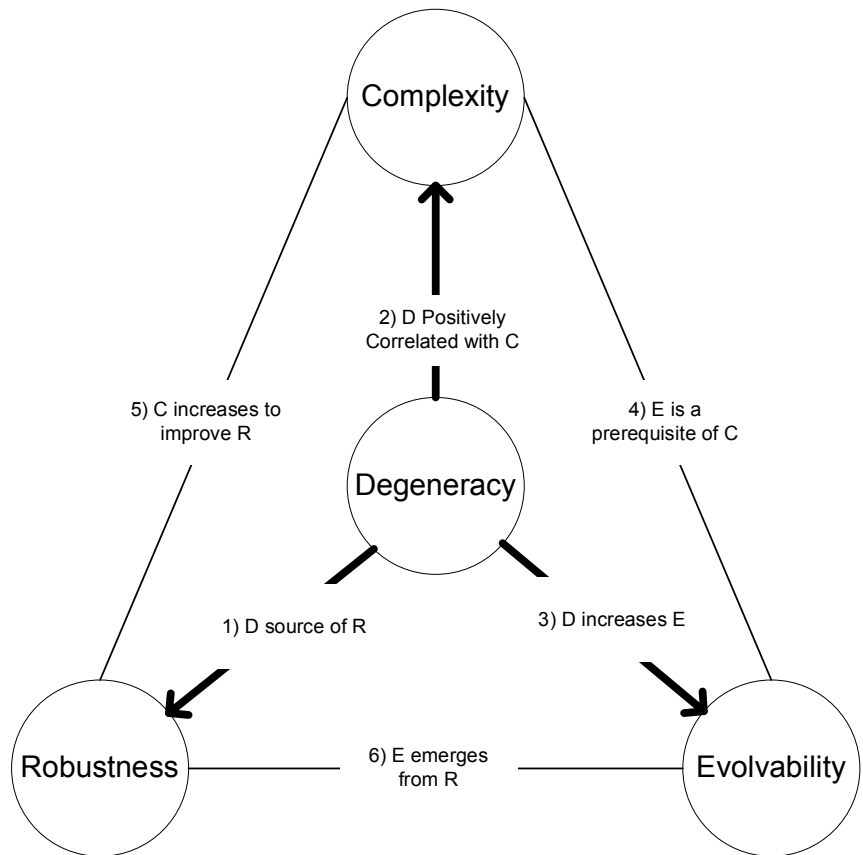

**Figure 1 high level illustration of the relationships between degeneracy, complexity, robustness, and evolvability. The numbers in column one of Table 1 correspond with the abbreviated descriptions shown here. This diagram is reproduced with permission from [3].**

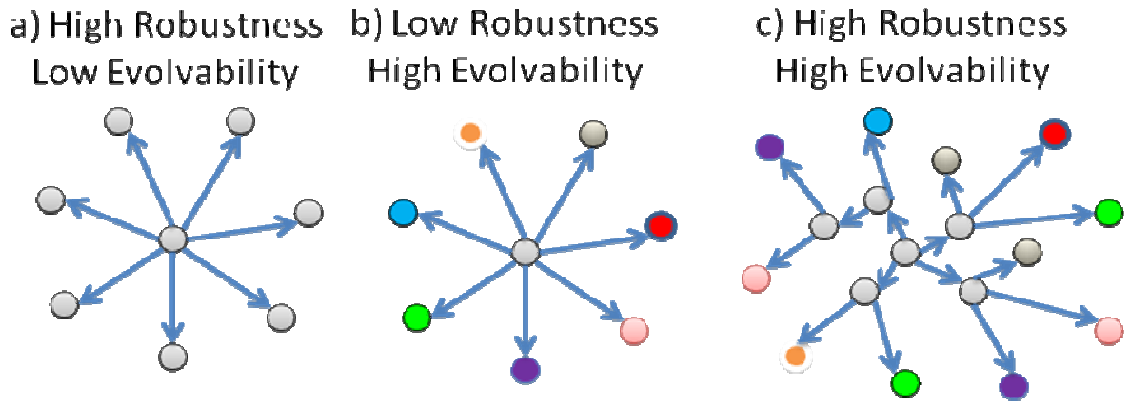

**Figure 2: The conflicting properties of robustness and evolvability and their proposed resolution. A system (central node) is exposed to changing conditions (peripheral nodes). Robustness of a function requires minimal variation in the function (panel a) while the discover of new functions requires the testing of a large number of functional variants (panel b). The existence of a neutral network may allow for both requirements to be met (panel c). In the context of a fitness landscape, movement along edges of each graph would reflect changes in genotype while changes in color would reflect changes in phenotype.**

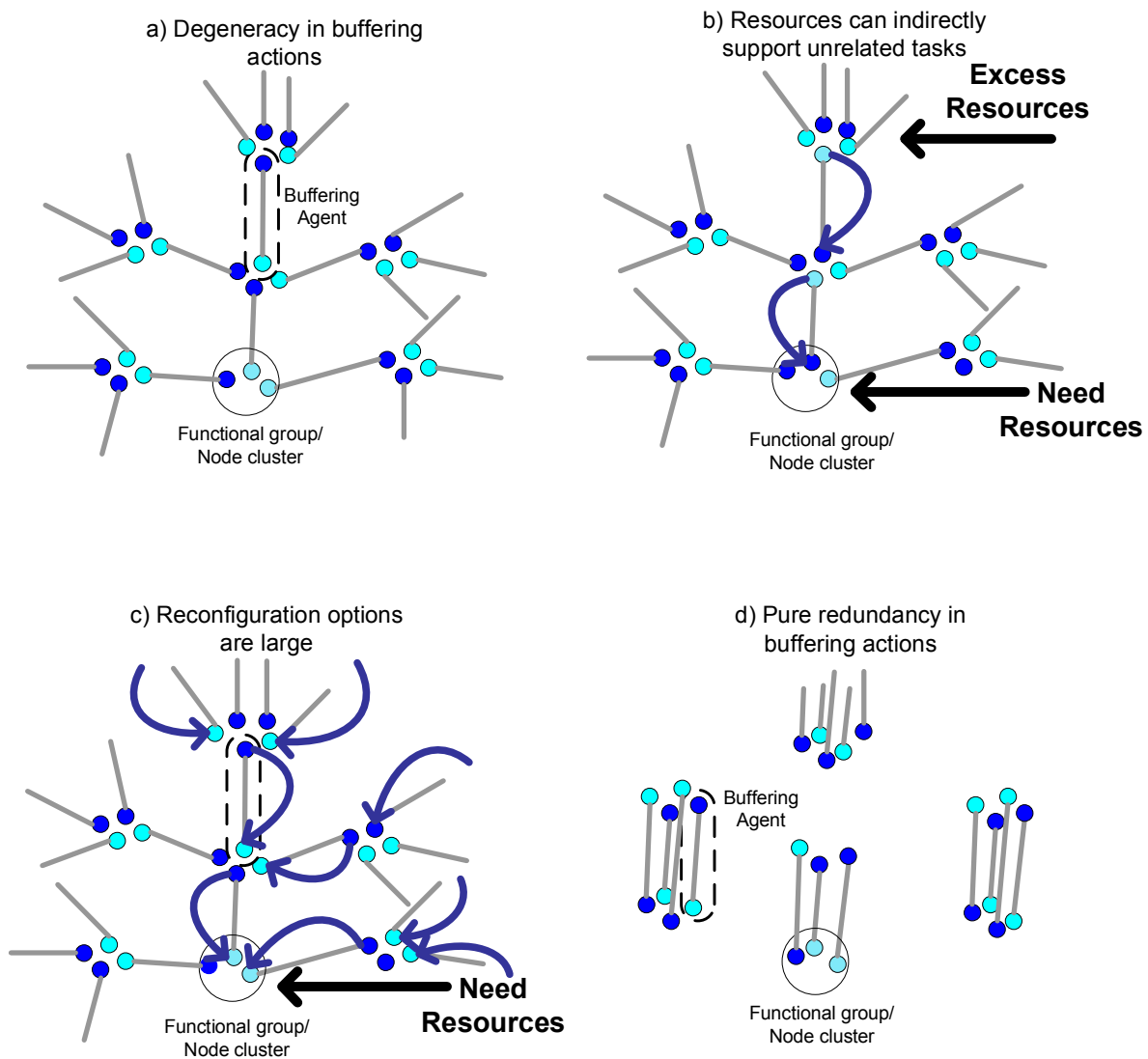

**Figure 3 Illustration of how distributed robustness can be achieved in degenerate systems (panels a-c) and why it is not possible in purely redundant systems (panel d). Nodes describe tasks, dark nodes are active tasks. In**

principle, agents can perform two distinct tasks but are able to perform only one task at a time. Panels a and d are reproduced with permission from [3].